\documentstyle[aps]{revtex}
\begin{document}

\title {Hysteresis and Avalanches in Two Dimensional Foam Rheology
Simulations} 
\author {Yi Jiang$^1$\cite{author1}, Pieter J. Swart$^1$, Avadh Saxena$^1$,
  Marius Asipauskas$^2$, James A. Glazier$^2$\\  
$^1$Theoretical Division, Los Alamos National Laboratory, Los Alamos,
NM 87545 \\ $^2$Department of Physics, University of Notre Dame, Notre Dame, IN 46556}   

\maketitle 

\begin{abstract}

Foams have unique rheological properties that range from solid-like to
fluid-like.  We study two-dimensional non-coarsening foams of
different disorder under shear in a Monte Carlo simulation, using a
driven large-Q Potts model.  Simulations of periodic shear on an
ordered foam show several different response regimes.  At small strain
amplitudes, bubbles deform and recover their shapes elastically, and
the macroscopic response is that of a linear elastic cellular
material.  For increasing strain amplitude, the energy-strain curve
starts to exhibit hysteresis before any topological rearrangements
occur, indicating a macroscopic viscoelastic response.  When the
applied strain amplitude exceeds a critical value, the yield strain,
topological rearrangements (T1 events) occur, the foam starts to flow,
and we observe macroscopic irreversibility.  We find that the dynamics
of topological rearrangements depend sensitively on the structural
disorder.  Structural disorder decreases the yield strain;
sufficiently high disorder changes the macroscopic response of a foam
from a viscoelastic solid to a viscoelastic fluid.  This wide-ranging 
dynamical response and the associated history effects of foams result
from avalanche-like T1 events.  The spatio-temporal statistics of T1
events T1 do not display long-range correlations for ordered foams or
at low shear rates, consistent with experimental observations.  As the
shear rate or structural disorder increases, the topological events
become more correlated and their power spectra change from that of
white noise toward $1/f$ noise.  Intriguingly, the power spectra of
the total stored energy also exhibit this $1/f$ trend.

\end{abstract}

\vskip 2\baselineskip

%\pacs{PACS numbers: 83.70.Hq, 82.70.Rr, 02.70.Lq, 46.60.Cn}
\vskip 2\baselineskip

\section{Introduction} 

In addition to their wide-spread industrial importance
\cite{kraynik88}, foams provide significant clues to the rheology of
other complex fluids, such as emulsions, colloids and polymer melts,
because we can observe their structures directly. 
The topological structures and the dynamics studied here also occur in other cellular materials, such as biological tissues and 
polycrystalline alloys.  One of the most remarkable and
technologically relevant features of foams is the range of mechanical
properties that arises from their structure.  For sufficiently small
stress, foams behave like a solid and are capable of supporting static
shear stress.  For large stress, foams flow and deform arbitrarily
like a fluid.  However, we do not yet fully understand the
relationship between the macroscopic flow properties of foams and
their microscopic details, {\it e.g.\/} liquid properties, topological
rearrangements of individual bubbles and structural disorder.
Constructing a full multiscale theory of foam rheology is challenging.
Foams display multiple length scales with many competing time scales,
memory effects ({\it e.g.\/} the hysteresis discussed in Sec. IV below),
and slow aging punctuated by intermittent bursts of activity ({\it
  e.g.\/} the avalanches of T1 events discussed in Sec. V below), all of
which severely limit their predictability and control.  These problems
are intriguing both from an applied and from a fundamental perspective
--- they provide beautiful concrete examples of multiscale materials,
where structure and ordering at the microscale, accompanied by fast
and slow time scales, can lead to a highly nonlinear macroscopic
response.  Here we study the relation between the microscopic
topological events and the macroscopic response in two-dimensional
non-coarsening foams using a driven extended large-Q Potts model. 

In foams, a small volume fraction of fluid forms a continuous network
separating gas bubbles \cite{kraynik88}.  The bubble shapes can vary
from spherical to polyhedral, forming a complex geometrical structure 
insensitive to details of the liquid composition or the average bubble
size \cite{gopal_durian95}.  Because of the complexity of describing
the network of films and vertices in three-dimensional foams, most
studies have been two-dimensional.  In two-dimensional foams free of
stress, all vertices are three-fold and the walls connecting them meet
at $120^\circ$ angles.  Minimization of the total bubble wall
length dictates that a pair of three-fold vertices is energetically
more favorable than a four-fold vertex.  Therefore, topology and
dynamics are intimately related, with the dominance of three-fold
vertices resulting from considerations of structural stability in the
presence of surface tension.  When shear stress is present, a pair of
adjacent bubbles can be squeezed apart by another pair (Fig. 1),
leading to a T1 switching event \cite{weaire_rivier84}.
This local but abrupt topological change results in bubble-complexes
rearranging from one metastable configuration to another.  The
resulting macroscopic dynamics is highly nonlinear and complex,
involving large local motions that depend on structures at the bubble
scale.  The spatio-temporal statistics of T1 events is fundamental to
the plastic yielding of two-dimensional liquid foams. 

The nonlinear and collective nature of bubble rearrangement dynamics
have made analytical studies difficult, except under rather special
assumptions.  Computer simulations can therefore provide important
insights into the full range of foam behavior.  Previous studies in
this field can be categorized through their use of constitutive,
vertex, center or bubble models. 

The constitutive models have evolved from the ideas of Prud'homme
\cite{prudhomme81} and Princen \cite{princen83}.  They modeled foam as
a two-dimensional periodic array of hexagonal bubbles where T1 events
occur instantaneously and simultaneously for the entire foam.  Khan
and Armstrong \cite{khan} further developed the model to calculate the
detailed force balance at the films and vertices, and studied the
stress-strain relationships as a function of hexagon orientation,
liquid viscosity and liquid fraction.  Reinelt and Kraynik
\cite{reinelt_kraynik90} extended the same model to study a
polydisperse hexagonal foam and derived explicit relations between
stress and strain tensors.  While analytical calculations exist only 
for periodic structures or for linear response, foams are naturally
disordered with an inherent nonlinear response.  Treating the foam as
a collection of interacting vertices, vertex models studied the
effect of stress on structure and the propagation of defects in foams
with zero liquid fraction ({\it i.e.\/} dry foam)
\cite{weaire_rivier84}.  Okuzono and Kawasaki
\cite{okuzono_kawasaki95} studied the effect of finite shear rate by
including the force on each vertex, a term which depends on the local
motion and is based on the work of Schwartz and Princen
\cite{schwartz87}.  They predicted avalanche-like rearrangements in a
slowly driven foam, with a power law distribution of avalanche
size {\it vs.\/} energy release, characteristic of self-organized criticality.
Durian's \cite{durian95,durian97} ``bubble'' model, treating bubbles  
as disks connected by elastic springs, measured foam's linear rheological
properties as a function of polydispersity and liquid fraction.  He
found similar distributions for the avalanche-like rearrangements with
a high frequency cutoff.  Weaire {\it et al.\/}~\cite{weaire_bolton92},
using a center model based on Voronoi construction from the bubble
centers, applied extensional deformation and bulk shear to a
two-dimensional foam.  They concluded that avalanche-like
rearrangements are possible only for wet foams, and that topological
rearrangements can induce ordering in a disordered foam.  A review by
Weaire and Fortes \cite{weaire_forte94} includes some computer models
of the mechanical and rheological properties of liquid and solid
foams.  However, few models have attempted to relate the structural
disorder and configuration energy to foam rheology.  Only recently,
Sollich {\it et al.\/} \cite{sollich97}, studying mechanisms for storing
and dissipating energy, emphasized the role of both structural
disorder and metastability in the rheology of soft glassy materials,
including foams.  Langer and Liu~\cite{langer97}, using a bubble model
similar to Durian's, found that the randomness of foam packing has a
strong effect on the linear shear response of a foam.  One of the
goals of our study is to quantify the extent of metastability by
measuring hysteresis, and relate the macroscopic mechanical response
to microscopic bubble structures.  

Experiments have measured the macroscopic mechanical properties of
three-dimensional foams.  But due to the difficulty of direct
visualization in three-dimensional foams, no detailed studies of
rearrangements exist.  Khan {\it et al.\/}~\cite{khan88}
applied bulk shear to a foam trapped between two parallel plates and
measured the stress-strain response, as well as the yield strain
as a function of liquid fraction.  Princen and Kiss
\cite{princen_kiss86}, applying shear in a concentric cylinder
viscometer ({\it i.e.\/} boundary shear), determined the yield stress
and shear viscosity of highly concentrated water/oil emulsions.
Recently, with the help of diffusing wave spectroscopy (DWS),
experiments by Gopal and Durian on three-dimensional shaving creams
showed that the rate of rearrangements is proportional to the strain
rate, and that the rearrangements are spatially and temporally
uncorrelated \cite{gopal_durian95}; H\"ohler {\it et
  al.\/}~\cite{hohler97} found that under periodic boundary shear, foam
rearrangements cross from a linear to nonlinear regime; H\'ebraud {\it
  et al.\/}~\cite{hebraud97} in 
a similar experiment on concentrated emulsions, found that some bubbles
follow reversible trajectories while others follow irreversible
chaotic trajectories.  However, none of these experiments has
directly observed changes in bubble topology.  Dennin and
Knobler \cite{dennin_knobler97} performed a bulk shear experiment on a
monolayer (2D) Langmuir foam and counted the number of bubble
side-swapping events.  Unfortunately, limited statistics rendered
their results difficult to interpret.   

In an attempt to reconcile the different predictions of different models and
experiments, we use a Monte Carlo model, the extended large-Q Potts
model, to study foam rheology.  The large-Q Potts model has
successfully modeled foam structure, coarsening and drainage
\cite{glazier90,jiang96}, capturing the physics of foams more
realistically than other models.  Here we extend the model to
include the application of shear to study the mechanical response 
of two-dimensional foams under stress.  

This paper is organized as follows: Sec. II presents our  
large-Q Potts model; Sec. III contains a description of simulation
details; Sec. IV presents results on hysteresis; Sec. V 
discusses the dynamics and statistics of T1 events; Sec. VI discusses
structural disorder and Sec. VII contains the conclusions.

\section{Model} 

The great advantage of our extended large-Q Potts model is its
simplicity.  The model is ``realistic'' in that the position and
diffusion of the walls determine the dynamics, as they do in real
foams and concentrated emulsions.  Previous models
\cite{okuzono_kawasaki95,durian97,weaire_bolton92} were based on
different special assumptions about the energy dissipation.  Since the
energy dissipation is poorly understood and also hard to measure in
experiments, the exact ranges of validity for these models are not
clear.  Not surprisingly, these models lead to conflicting
predictions, {\it e.g.\/} for the distribution of avalanche-like
rearrangements (Sec. V).  None of these models alone captures the full
complexity of real foams. 

The extended large-Q Potts model, where bubbles have geometric
properties as well as surface properties, is not based on any {\it a priori\/} 
energy dissipation assumption.  In addition, it has the advantage of
simultaneously incorporating many interactions, including temperature
effects, for foams with arbitrary disorder and liquid content
\cite{note1}. 

Both the film surface properties and the geometry of bubbles are
fundamental to understanding foam flow.  The contact angle of walls
between vertices indicates whether the structure is at equilibrium,
corresponding to minimizing the surface energy.  In a real evolving
pattern, the equilibrium contact angle occurs only for slow movements
during which the vertices remain adiabatically equilibrated.  Whenever a
topological rearrangement (a T1 event) of the pattern occurs, the
contact angles can be far from their equilibrium values.  The walls
then adjust rapidly, at a relaxation rate depending on the effective
foam viscosity, to re-establish equilibrium.  The same holds true for
the other possible topological change, the disappearance of a bubble,
a T2 event \cite{weaire_rivier84}.  However, disappearance only occurs
in foams that do not conserve bubble number and area, which we do not
consider in this study.  A difficulty in two-dimensional foams is that
the effective viscosity depends primarily on the drag between the
Plateau borders and the top/bottom surfaces of the container, not the
liquid viscosity.  Container chemistry, surfactant properties, and
foam wetness all change the effective viscosity.  Thus even in
experiments, the effective viscosity is not equivalent to the liquid
viscosity and not possible to derive from liquid viscosity.  We define
the equilibrium contact angle so that any infinitesimal displacement
of the vertex causes a second-order variation of the surface energy,
while during a T1 event the energy must vary macroscopically over a
small but finite coherence length, typically the rigidity length of a
bubble.  In our simulations, a bubble under stress can be stretched or
compressed up to $60\%$ of its original length, while conserving its
area. 

In a center model based on the Voronoi construction (see, {\it e.g.\/},
\cite{weaire_bolton92}), the coherence length of a bubble is 
comparable to its diameter.  Contact angles are given correctly at
equilibrium but approach and remain near $90^\circ$ during a T1 event,
since the centers are essentially uninfluenced by topological details
such as the difference between a four-fold vertex and a pair of
three-fold vertices.  

In a vertex model (see, {\it e.g.\/},~\cite{okuzono_kawasaki95}), the walls
connecting the vertices adiabatically follow an out-of-equilibrium,
slowly-relaxing vertex.  In such a model, the walls are constrained
to be straight and vertices typically have arbitrary angles.  In
essence, the deviation of the vertex angles from the equilibrium value
represents the integrated curvature of the bubble walls.  Because of 
their unphysical representation of contact angles, pure vertex models 
with straight walls cannot handle T1 events correctly.

The extended large-Q Potts model avoids these limitations:
walls are free to fluctuate, which is not true in vertex models; and the 
contact angles during a T1 event are correct, which is not true in
center models.  A further advantage of our extended large-Q Potts
model is that it allows direct measurement of T1 events.  The
other models cannot directly count T1 events.
Instead, they quantify rearrangement events by their associated
decreases in energy.  We will discuss later this energy decrease
is not always directly proportional to the number of T1 events.  Our
model therefore delivers accurate information about individual T1
events as well as the averaged macroscopic measures such as total
bubble wall length, thereby allowing new insights into the connection
between microscopic foam structure and macroscopic mechanical
response.  

Before describing the details of the Potts model, we should first
mention its major limitations.  Viscosity is one of the basic physical  
properties of foams, but it is not easily specified {\it a priori\/} in
the Potts model.  Although we can extract the effective viscosity and
the viscoelasticity of foams from simulations, we lack a clear
quantitative description of the foam viscosity in Potts model  
simulations and how it relates to the effective and liquid viscosities
of a two-dimensional foam.  However, our ignorance about simulation
viscosity is equivalent to our ignorance about experimental
two-dimensional foam viscosity.  Quantitative experiments will help to 
separate the roles of the Plateau borders, fluid viscosity, and
topological rearrangements in determining the effective foam
viscosity.  A second possible limitation is the size effect due to
lattice discretization.  We show in Sec. III that this problem does
not invalidate our simulations.  A third drawback is that the Monte
Carlo algorithm results in uncertainties in the relative timing of
events on the order of a few percent of a Monte Carlo step.  While
this uncertainty is insignificant for well separated events, it can
change the measured interval between frequent events.   

The extended large-Q Potts model treats foams as spins on a lattice.
Each lattice site $i=(x_i, y_i)$ has an integer ``spin''  $\sigma_{i}$
chosen from $\{1, \ldots , Q\}$.  Domains of like spins form bubbles,
while links between different spins define the bubble walls (films).  
Thus each spin merely acts as a label for a particular bubble.
The surface energy resides on the bubble walls only.  Since the
present study focuses on shear-driven 
topological rearrangements over many loading cycles, we prohibit foam
coarsening by applying an area constraint on individual bubbles.  In
practical applications, foam deformation and rearrangement under
stress is often much faster than gas diffusion through the walls, so
neglecting coarsening is reasonable.  The Potts Hamiltonian, the total
energy of the foam, includes the surface energy and the elastic bulk
energy:  

\begin{equation}
{\cal H} = \sum_{ij}{\cal J}_{ij} (1-\delta _{\sigma_{i} \sigma_j}) 
         + \Gamma \sum_n (a_n-A_n)^2 ,
\end{equation}

\noindent where ${\cal J}_{ij}$ is the coupling strength between
neighboring spins $\sigma_i$ and $\sigma_j$, summed over the 
entire lattice.  The first term gives the total surface energy.  The
second term is the area constraint which prevents coarsening.  The
strength of the constraint ($\Gamma$) is inversely proportional to the 
gas compressibility; $a_n$ is the area of the $n$th bubble and $A_n$
its corresponding area under zero applied stress.  We can include 
coarsening by setting $\Gamma$ to zero.  

We extend the Hamiltonian to include shear: 

\begin{equation}
{\cal H}' = {\cal H}  
+ \sum_i \gamma (y_i, t) x_i (1-\delta_{\sigma_i \sigma_j}) .
\end{equation}

\noindent The new term corresponds to applying shear strain (a detailed
explanation follows below) to the wall between neighboring bubbles
$\sigma_i$ and $\sigma_j$, with $\gamma$ corresponding to the strain 
field, $(x_i, y_i)$ to the coordinate of spin $\sigma_i$ and $(1,0)$
is the direction of the strain. 

The system evolves using Monte Carlo dynamics.  Our algorithm 
differs from the standard Metropolis algorithm: we choose a spin at 
random, but {\em only\/} reassign it if it is at a bubble wall and then
{\em only\/} to one of its unlike neighbors.  The probability 
of accepting the trial reassignment follows the Boltzmann distribution,
namely:  

\begin{equation}
  P \sim \left\{ \begin{array}{lll}
      1  & \mbox{$\Delta {\cal H}'<0$} \\  
      \exp (-\Delta {\cal H}' /T) & \mbox{$\Delta {\cal H}' \geq 0$}
    \end{array}
  \right. , 
\end{equation}

\noindent where $\Delta {\cal H}'$ is the change in ${\cal H}'$ due to
a trial spin flip, and $T$ is temperature.  Time is measured in units
of Monte Carlo steps (MCS), where one MCS consists of as many spin
trials as there are lattice sites.  This algorithm reproduces the same 
scaling as classic Monte Carlo methods in simulations of foam
coarsening, but significantly reduces the simulation time~\cite{potts}. 

The second term in ${\cal H}'$ biases the probability of spin
reassignment in the direction of increasing $x_i$ (if $\gamma<0$) or
decreasing $x_i$ (if $\gamma>0$).  From dimensional analysis of ${\cal
  H'}$, $\gamma$ has units of force, but we can interpret it as
the strain field for the following reason:  In the Potts model
a bubble wall segment moves at a speed proportional to the
reassignment probability $P$; in this case,    

\begin{equation}
v \propto \sqrt{\gamma}P , 
\end{equation}

\noindent where the prefactor follows from dimensional analysis.  This
shear term effectively enforces a velocity $v$ at the bubble  
walls, therefore it imposes a strain rate on the foam.  The strain 
$\epsilon(t)$ is then proportional to a time integral of $v$,

\begin{equation}
\epsilon \propto \int_0^t \sqrt{\gamma(t')}P dt'. 
\end{equation}

If we limit the application of this term to the boundaries of the foam, 
we impose a boundary shear, equivalent to moving the boundary of the
foam with no-slip between bubbles touching the boundary and the
boundary, {\it i.e.\/},    

\begin{equation}
  \gamma= \left \{ \begin{array}{ll}
      \gamma_0 G(t) & y_i=y_{{\rm min}} \\
      -\gamma_0 G(t) & y_i=y_{{\rm max}} \\
      0   & {\textstyle otherwise }
    \end{array}
  \right. ,
\end{equation}

\noindent where $\gamma_0$ is the amplitude of the strain field and $G(t)$
is a normalized function of time.  On the other hand, 

\begin{equation}
  \gamma= \beta y_i G(t) ,
\end{equation}

\noindent with $y_i$ between $y_{{\rm min}}$ and $y_{{\rm max}}$,
corresponds to applying bulk shear with the strain rate varying
linearly as a function of position in the foam.  The gradient of 
strain rate is the shear rate, $\beta$.  The corresponding experiment
would be similar to Dennin and Knobler's monolayer Langmuir foam
experiment \cite{dennin_knobler97}: a monolayer foam (2D) on the
surface of a liquid is sheared in a concentric Couette cell, with
no-slip conditions between the bubbles and the container surface. 
In all our studies we use $G(t)=1$ for steady shear, and
$G(t)=\sin(\omega t)$ for periodic shear.  Since for steady shear, the
strain is a constant times time, or $\sqrt{\gamma}Pt$, plotting with respect 
to time is equivalent to plotting with respect to strain.

Note that our driving in the Potts model differs from that in driven 
spin systems, for which a large body of literature addresses the
dynamic phase transition as a function of driving frequency and
amplitude \cite{acharyya95}.  Our driving term acts on the bubble
walls (domain boundaries) only, while in driven spin systems {\it
  e.g.\/} the kinetic Ising model, all spins couple to the driving field.
The resulting dynamics differ greatly. 

\section{Simulation details}  

Experimental observations show that the mechanical responses of a foam,
including the yield strain, the elastic moduli, and the topological
rearrangements, are sensitive to the liquid volume fraction 
\cite{weaire_pittet93}.  In particular, the simulations of both Durian
\cite{durian97} and Weaire {\it et al.\/} \cite{weaire_bolton92} showed
a critical liquid fraction at which a foam undergoes a ``melting
transition.''  Although different liquid content and drainage effects
can be readily incorporated in the Potts model \cite{jiang96}, we
focus on the dynamics of topological rearrangements and do not
consider liquid fraction dependence of flow behavior, {\it i.e.\/} we
assume the dry 
foam limit in this study.  Also, we ignore gas diffusion across the
walls, assuming that bubble deformation and rearrangement are much
faster than coarsening.   

The definition of time (Monte Carlo steps or MCS) is not directly
related to real time, but we have made choices to ensure that we do
not under-resolve events.  A shear cycle in the periodic shear case
takes about 4000 MCS.  In our simulations, a single deformed bubble
recovers on a timescale of a few MCS while the relaxation of a cluster
of deformed bubbles takes a much longer time, on the order of 10 to 100
MCS.  A T1 event by definition takes one MCS (the short life of a
four-fold vertex), but the viscous relaxation has to average over at
least the four bubbles involved in the T1 event, and thus lasts much
longer.  

We used periodic boundary conditions in the $x$ direction, to mitigate
finite size effects.  For ordered foams under boundary shear, we
used a $400\times100$ lattice with each bubble containing $20\times20$
lattice sites; for ordered foams under bulk shear, we used a
$256\times256$ lattice with $16\times16$ sites for each bubble.  When
unstressed, all the bubbles are hexagons, except for those truncated
bubbles touching the top and bottom boundaries.  In the case of
disordered foams, we used a $256\times256$ lattice with various area
distributions.  We have also performed simulations using a lattice of
size $1024\times1024$ with $64\times64$ bubbles and a lattice of size
$1024\times1024$ with $16\times16$ bubbles.  The results did not 
appear to differ qualitatively.  A $16\times16$ bubble has a side
length around 10 lattice sites, so its smallest resolvable tilt angle is
approximately $arctan(1/10) \sim 5.7^\circ$.  Had lattice effects been
a problem, we would have expected a significant difference in the
simulations with bubbles of size $64\time64$, where the smallest angle
is about four times smaller.  But increasing the simulation size from
$16^2$ to $64^2$ did not lead to significant changes in the
quantities we measured.  Thus, we used bubbles of size $16^2$ in
all the simulations reported in this paper. 

Lattice anisotropy can induce artificial energy barriers in lattice  
simulations.  All our runs use a fourth-nearest neighbor interaction
on a square lattice, which has a lattice anisotropy of $1.03$, very
close to the isotropic situation (lattice anisotropy of 1). 

Standard quantitative measures of cellular patterns are the
topological distributions and correlations, area distributions, and
wall lengths --- all quantities that in principle can be measured in
experiments.  Since the areas are constrained, the evolution of the area
distribution is not useful.  We define the topological distribution
$\rho(n)$ as the probability that a bubble has $n$ sides; its $m$-th
moments are $\mu_m \equiv \sum_n \rho(n) (n - \langle n \rangle)^m $.
The area distribution $\rho(a)$ and its second moment $\mu_2(a)$ are
defined in a similar fashion for the bubble areas.  We use a variety
of disordered foams with different distributions, as characterized
by their $\mu_2(n)$ and $\mu_2(a)$.  

In practice, we generate the initial configuration by partitioning the
lattice into equal-sized square domains, each containing $16\times16$
lattice sites.  The squares alternate offsets in every other row, so the
pattern resembles a brick wall arranged in common bond.  We then run
the simulation with area constraints, but without strain, at finite
temperature for a few Monte Carlo steps, and then decrease the
temperature to zero and let the pattern relax.  The minimization of
total surface energy (and hence the total bubble wall length) results
in a hexagonal pattern, the initial configuration for the ordered
foam.  For disordered initial configurations, we continue to evolve
the hexagonal pattern without area constraints at finite temperature
so that the bubbles coarsen.  We monitor $\mu_2(n)$ of the evolving
pattern, and stop the evolution at any desired distribution or degree
of structural disorder.  Then we relax the patterns at zero
temperature with area constraints to guarantee that they have
equilibrated, {\it i.e.\/} without added external strain or stress the
bubbles would not deform or rearrange.   

For all our simulations, $\Gamma=1$ (which is sufficiently large to
enforce air incompressibility in bubbles) and ${\cal J}_{ij} =3$
(except when we vary the coupling strength to change the effective
viscosity of the foam).  Most of the simulations shown in this paper are
run at zero temperature except when we study temperature effects
on hysteresis, because the data are less noisy and easier to
interpret.  A finite but low temperature speeds the simulations, but
does not appear to change the results qualitatively.   

The number of sides of a bubble is defined by its number of different 
neighbors.  During each simulation, we keep a list of neighbors for
each bubble.  A change in the neighbor list indicates a topological
change which, since bubbles do not disappear, has to be a T1 event.

\section{Hysteresis} 

We can view foam flow as a collective rearrangement of bubbles from
one metastable configuration to another.  We investigate the
configurational metastability by studying hysteresis of the
macroscopic response.  

Hysteresis is the phenomenon in which the macroscopic state of a
system does not reversibly follow changes in an external parameter, 
resulting in a memory effect.  Hysteresis commonly appears in systems
with many metastable states due to (but not limited to) interfacial
phenomena or domain dynamics.  The classic example of the former is
that the contact angle between a liquid and a solid surface depends on
whether the front is advancing or retreating.  The classic example of
the latter is ferromagnetic hysteresis, in which the magnetization
lags behind the change in applied magnetic field.  In cellular
materials, including foams, hysteresis can have multiple microscopic
origins, including stick-slip interfacial and vertex motion, local
symmetry-breaking bubble rearrangement (T1 events), and the nucleation
of new and annihilation of old cells.  In all of these, noise and
disorder play an intrinsic role in selecting among the many possible
metastable states arising when the foam is driven away from
equilibrium.  By focusing on non-coarsening foams, we rule out
nucleation and annihilation as sources for hysteresis.  Our foam is
therefore an ideal testing ground for improving our understanding of
hysteresis as it arises from local rearrangements and interfacial dynamics.

In accordance with \cite{okuzono_kawasaki95}, we define the quantity:

\begin{equation}
  \phi \equiv \sum_{i,j} \theta(1- \delta_{\sigma_i,\sigma_j}) ,  
\end{equation}

\noindent as the total stored elastic energy.  Here sites $i, j$
are neighbors, summation is over the whole lattice, and $\theta$ is the
wall thickness that we choose to be $1$ in all our simulations (dry
foam limit).  Thus $\phi$ gives essentially the total bubble wall length, 
which differs by a constant, namely the surface tension, from the
total surface energy.  In zero temperature simulations, the area
constraint is almost always satisfied so that small fluctuations in
areas contribute  only $10^{-3}$ of the total energy.  Thus we can
neglect the elastic bulk energy of the bubbles, and assume that the
total foam energy resides on the bubble walls only, {\it i.e.\/} all forces
concentrate at the bubble walls.  We can calculate values of the averaged 
stress by taking numerical derivatives of the total surface energy
with respect to strain \cite{weaire_bolton92}.  However, the
calculation via derivatives is not suitable for foams undergoing many
topological changes, since the stored elastic energy changes discontinuously
when topological rearrangements occur.  The alternative is to
calculate stress directly, as given in \cite{weaire_hutzler95}, by the
sum of forces acting on the bubble walls, which locally is
proportional to the wall length change of a bubble.  Because forces on
the bubble walls in Potts model foams are not well characterized, we
limit our discussions to energy-strain relationships.  The more
rigorous definition of strain involves the definition of a mesoscopic
lengthscale corresponding to a cluster of bubbles, over which the
effects of bubble wall orientation and bubble deformation can be
averaged.  In \cite{okuzono_kawasaki95}, the average stress tensor,
defined as $\sigma = 1/A \sum_{\langle{i,j}\rangle} |r_{ij}| \hat
r_{ij}\hat r_{ij}$, with $A$ the total area of the foam and $r_{ij}$
the distance between two neighboring vertices, is directly related to
$\phi$ via $\phi = Tr(\sigma)$.  Hereafter, we present our $\phi$ data 
as $\phi(t)/\phi(0)$ to scale out differences due to initial
configurations.   

\subsection{Hysteresis in ordered foams}

The simplest perturbation which induces topological rearrangements is
boundary shear on an ordered foam.  In this case we can confine
the deformation to the bubbles touching the moving boundaries, and
easily locate all the T1 events.  As the applied boundary shear
increases, the bubbles touching the boundaries distort, giving rise to
a stored elastic energy.  We show snapshots of the pattern in Fig. 2(a).  
When a pair of vertices come together to form a four-fold
vertex, the numbers of sides changes for the cluster of bubbles
involved.  Different shades of gray in Fig. 2(a) reflect the
topologies of the bubbles.  Note that a five-sided (dark grey) and a
seven-sided (light grey) bubble always appear in pairs except during
the short lifetime of a four-fold vertex (when the number of sides
are ambiguous because of the discrete lattice).  Once the strain
exceeds a critical value, the yield strain, all the bubbles touching
the moving boundaries undergo almost simultaneous rearrangements,
thereby releasing stress.  The stored elastic energy, $\phi$, increases
with time when the bubbles deform, then decreases rapidly when the
bubbles rearrange.  Stress accumulates only in the two 
boundary layers of bubbles, and never propagates into the interior of
the foam.  The whole process repeats periodically, due to the periodic
bubble structure, as shown in Fig. 2(b), the energy-strain plot (as
mentioned at the end of Sec. II, for steady shear, plotting time is
equivalent to plotting strain).  This result corresponds to the
mechanical response obtained in the model of Khan {\it et al.\/} with
periodic hexagonal bubbles oriented at zero degrees with respect to
applied strain \cite{khan}.    

When applying periodic shear $\gamma(t)=\gamma_0 \sin(\omega t)$, we keep
the period $2 \pi / \omega$ fixed and vary the amplitude, $\gamma_0$.
Under sinusoidal periodic shear, we observe three types of behavior.
When the strain amplitude is small, bubbles deform and recover their
shapes elastically when stress is released.  No topological
rearrangement occurs and the energy-strain plot is linear, corresponding
to an elastic response \cite{note2}.  This result agrees perfectly
with the experimental result of DWS in \cite{hohler97}.  As
the strain amplitude increases, the energy-strain curve begins to
exhibit a small butterfly-shaped hysteresis loop before any
topological rearrangements occur, indicating a macroscopic
viscoelastic response.  If we keep increasing the strain amplitude,
the hysteresis loop increases in size.  When the applied strain
amplitude exceeds a critical value, T1 events start occurring, and the foam
starts to flow, which leads to a further change in the shape of the
hysteresis loop.  Even larger strain amplitude introduces more T1
events per period, and adds small loops to the ``wings'' of the
hysteresis loop.  Figure 3(a) shows the smooth transition between the
three types of hysteresis in the energy-strain curve.  

We can adjust the viscosity of the bubble walls by changing the
coupling strength ${\cal J}_{ij}$.  Smaller coupling strength
corresponds to lower viscosity.  Similar transitions from elastic, to
viscoelastic to fluid-like flow behavior occur for progressively lower
values of coupling strength, shown in Fig. 3(b).  The phase diagram in
Fig. 3(c) summarizes $44$ different simulations and shows the elastic,
viscoelastic and fluid-like behavior (as derived from the hysteretic
response) as a function of the coupling strengths ${\cal J}_{ij}$
({\it i.e.\/} viscosity) and strain amplitudes $\gamma_0$.  A striking
feature is that the boundaries between these regimes appear to be
linear.  Figure 3(d) shows the effect of finite temperature on the
energy-strain curves.  With progressively increasing temperature,
noise becomes more dominant and eventually destroys the hysteresis
loop.  This result implies diminished metastability at finite
temperature.  However it does not seem to change the trend in
mechanical response.  

A more conventional experiment is the application of bulk shear
\cite{okuzono_kawasaki95,durian97,princen_kiss86,dennin_knobler97}, 
with the shear strain varying linearly as a function of the vertical
coordinate, from $\gamma_0$ at the top of the foam to  $-\gamma_0$ at
the bottom.  In our bulk shear simulations with an ordered foam, the
energy-strain relationship has two distinct behaviors depending on the
shear rate.  At small shear rates, a ``sliding plane'' develops in
the middle of the foam.  As shown in Fig. 4(a), non-hexagonal bubbles
appear only at the center plane.  The energy-strain curve, shown in 
Fig. 4(b), therefore, resembles that for boundary shear
on an ordered foam.  The energy curve in Fig. 4(b) also shows that the 
baseline of energy is larger and that the decrease in
amplitude of the energy due to T1s is smaller than in Fig. 2(b),
because bulk shear induces a more homogeneous distribution of 
distortion and thus of stored elastic energy.  Again the periodic
structure of the bubbles cause the periodicity of the curve,
reminiscent of the shear planes observed in metallic glasses in the
inhomogeneous flow regime, where stress-induced rearrangement causes 
plastic deformation \cite{spaepen97}.    

At high shear rates, the ensemble of T1 events no longer localize
in space [Fig. 5(a)].  Non-hexagonal bubbles appear throughout the foam.  The
energy-strain curve, shown in Fig. 5(b), is not periodic but rather
smooth; beyond the yield point, the bubbles constantly move
without settling into a metastable configuration and correspondingly,
the foam displays dynamically induced topological disorder. 
The transition between these two regimes, localized and nonlocalized
T1 events, occurs when the shear rate is in the range  
$1 \times 10^{-2} < |\beta| < 5 \times 10^{-2}$.  This transition can be 
understood if we look at the relaxation time scale of the foam.  Due
to surface viscous drag and geometric confinement of other bubbles,
the relaxation time for a deformed bubble in a foam is on the order of
$10$ MCS.  For a shear rate $\beta = 5 \times 10^{-2}$, $\beta ^{-1}$
is of the same order as the relaxation time.  Thus for shear rates
above the natural internal relaxation time scale, the macroscopic
response changes from jagged and piece-wise elastic to smooth and
viscous response, as observed in fingering experiments in
foams~\cite{park_durian94}.   

\subsection{Hysteresis in disordered foams}

In a disordered foam, bubbles touching the moving foam boundary have
different sizes.  Boundary strain causes different bubbles to undergo
T1 rearrangements at different times.  Stress no longer localizes in  
(sliding)  boundary layers, but propagates into the interior [Fig. 6(a)].  
The yield strain is much smaller.  When the size distribution of the
foam is broad, the linear elastic regime disappears, since even a small
strain may lead to topological rearrangements of small bubbles.
In other words, with increasing degree of disorder, the yield strain
decreases to zero and the foam changes from a viscoelastic solid to
a viscoelastic fluid.  We show an
example of such viscoelastic fluid behavior in Fig. 6(b) for a
random foam, which shows no energy accumulation, namely,
its yield strain is zero.  The foam deforms and yields like a fluid
upon application of the smallest strain.   

Under a periodic shear, the stored energy increases during an initial
transient period but reaches a steady state after a few periods of
loading.  Energy-strain plots show hysteresis due to topological
rearrangements  
similar to those in ordered foams, but as the degree of disorder
increases, the corresponding elastic regime shrinks and eventually
disappears.   

Rearrangement events in a disordered foam under bulk shear at a low
shear rate [snapshots shown in Fig. 7(a)] correspond to those in an
ordered foam at a high shear rate.  The rearrangements are discrete
and avalanche-like, resembling 
a stick-slip process, or adding sand slowly to a sandpile.  However,
at sufficiently high shear rate all the avalanches overlap and the
deformation and rearrangements are more homogeneous and continuous, as 
in a simple viscous liquid.  Figure 7(b) shows the typical energy-strain
curve of a disordered foam under steady bulk shear.

Note that in all our hysteresis plots, the energy-strain curves cross
at zero strain, indicating no residual stored energy at zero strain.
This crossing is an artifact of our definition of energy, which ignores
angular measures of distortion, {\it i.e.\/} the total bubble wall
length does not distinguish among the directions in which the bubbles
tilt.  A choice of stress definition which included angular
information would show some residual stress at zero strain, but would
not affect the results reported here.

\section{T1 avalanches} 

In both experiments \cite{glazier89,stavans89} and our simulations,
the contact angles of the vertices remain close to $120^\circ$ until
two vertices meet.  The applied strain rate determines the rate at
which vertices meet.  The resulting four-fold vertex rapidly splits 
into a vertex pair, recovering $120^\circ$ contact angles, at a rate 
determined by the viscosity.  This temporal asymmetry in the T1
event contributes to the hysteresis. 

In vertex model simulations, sudden releases of energy occur once the
applied shear exceeds the yield strain \cite{okuzono_kawasaki95}.  
The event size $n$, {\it i.e.\/} the energy release per event in the dry
foam limit, follows a power-law distribution: $\rho(n) \sim
n^{-\frac{3}{2}}$.  Durian \cite{durian97} found a similar power-law 
distribution in his bubble model, with an 
additional exponential cutoff for large events.  
Simulations of Weaire {\it et al.\/}
\cite{weaire_bolton92,weaire_hutzler95}, however, suggested that power-law 
behavior only appeared in the wet foam limit.  Experiments, on the
other hand, have never found system-wide events or long range
correlations among events \cite{gopal_durian95,dennin_knobler97}.  One
of our goals is to reconcile these different predictions.

These differences may result from the use of energy release as proxy for
topological changes, rather
than enumerating actual events, as well as the assumption of a linear
relation between jumps in the stored elastic energy and the number of
T1 events, namely, ${d\phi / dt} = cN $, where $N$ is the number of T1
events and $c$ is a constant.  A drastic drop in the total bubble wall
length indicates a large number of T1 events.  However, in a
disordered foam, all T1 events are not 
equal, since they do not all release the same amount of stored elastic
energy.  The energy released during a T1 event scales as the bubble
perimeter, {\it i.e.\/} smaller bubbles release less energy.
Hence smaller bubbles undergo more T1
events.  Moreover, a T1 event is not strictly local, but deforms its 
neighborhood over a certain finite range, as
demonstrated by T1 manipulations in magnetic fluid foam experiments
\cite{elias98}.  Therefore, the number of T1 events is not always
directly proportional to the decrease in total bubble wall length.  Thus
we cannot compare the energy dissipation and T1 events
directly.  Furthermore, the mechanisms of energy dissipation differ in
these models.  Kawasaki {\it et al.\/} \cite{okuzono_kawasaki95} included
the dissipation due to the flow of liquid out of the Plateau borders;
Durian \cite{durian97} considered only the viscous drag of the liquid,
while Weaire {\it et al.\/}~\cite{weaire_bolton92} modeled an
equilibrium calculation involving quasistatic steps in the strain that
do not involve any dissipation.  In our model, the evolution minimizes
the total free energy naturally.  To avoid ambiguities, we directly
count T1 events in addition to tracking energy. 

The avalanche-like nature of rearrangements appears in the sudden
decreases of the total elastic energy as a function of time.  Figure
4(b) shows the relation between energy and the number of T1 events in
an ordered foam under steady bulk shear for a small strain.  The
stored energy increases almost linearly until the yield strain is
reached.  The avalanches are well separated.  Every cluster of T1
events corresponds to a drastic decrease in the stress, and the
periodicity is due to the ordered structure of the foam.  At a higher
shear rate [Fig. 5(b)], the yield strain remains almost the same, but
the avalanches start to overlap and the energy curve becomes
smoother.  In the sandpile analogy, instead of adding sand grains one
at a time and waiting until one avalanche is over before dropping
another grain, the grains accumulate at a constant rate and the
avalanches, large and small, overlap one another.  A sufficiently
disordered foam may not have a yield strain [Fig. 6(b)]; T1 events
occur at the smallest strain.  The foam flows as a fluid without
going through an intermediate elastic regime.  

To study the correlation between T1 events, we consider the power
spectrum of $N(t)$, the number of T1 events at each time step,

\begin{equation}
p_N(f) = \int dt \int d\tau e^{-if \tau} N(t) N(t+\tau) ,
\end{equation}

\noindent where $f$ is the frequency with unit $MCS^{-1}$.  Figure
8(a) shows typical power spectra of the time series of T1 events in an
ordered foam under bulk shear.  At a shear rate $\beta=0.01$, the T1
events show no power law.  The peak at $\sim 10^{-3}$ is due
to the periodicity of bubble structure in an ordered foam when a
``sliding plane'' develops.  At shear rate $\beta=0.02$, the spectrum
resembles that of white noise.  As the shear rate increases to
$\beta=0.05$, the power spectrum develops a power law tail at the low
frequency end, with an exponent very close to 1.  In a disordered
foam, with increasing shear rate, the spectra for the T1 events
gradually change from completely uncorrelated white noise to $1/f$ at
higher shear rates.  By $1/f$, we mean any noise of power spectrum
$S(f) \sim f ^{-\alpha}$ where $0<\alpha<2$ or near 1, {\it i.e.\/}
intermediate between Brownian noise ($\alpha=2$) and white noise
($\alpha =0 $). 

These power spectra suggest that the experimental results for T1
events as reported in~\cite{gopal_durian95,dennin_knobler97}
correspond to a low shear rate, with no long-range correlation among 
T1 events.  Structural disorder introduces correlations among the
events.  Power-law avalanches do not occur in ordered hexagonal cells
at low shear rate, where rearrangements occur simultaneously.  At a
high shear rate, when the value of $\dot \gamma^{-1}$ is comparable to
the duration of rearrangement events, the bubbles move constantly.
The foam behaves viscously, since rearrangements are continuously
induced before bubbles can relax  into metastable configurations which 
can support stress elastically.  At these rates, even an initially
ordered structure behaves like a disordered one, as shear destroys its
symmetry and periodicity.

In a disordered foam, whenever one T1 event happens, the deformed bubbles
release energy by viscous dissipation and also transfer stress to their 
neighboring bubbles, which in turn are more likely to undergo a T1 switch.
Thus, T1 events become more correlated.  Shown in Fig. 8(b), the power
spectra change from that of white noise toward $1/f$ noise.   
When the first sufficiently large region to accumulate stress undergoes
T1 events, it releases stress and pushes most of the rest of the bubbles 
over the brink, causing an ``infinite avalanche'': some bubbles 
switch neighbors, triggering their neighbors to rearrange (and so on),
until a finite fraction of the foam has changed configuration, causing
a decrease in the total stored energy, mimicking the cooperative
dynamic events in a random field Ising model \cite{sethna93}. 
We never observe system-wide avalanches as claimed in the vertex
model simulations~\cite{okuzono_kawasaki95}, agreeing with Durian's
simulations~\cite{durian97} and Dennin {\it et al.\/}'s
experiments~\cite{dennin_knobler97}.  For even greater disorder,
the bubbles essentially rearrange
independently, provided spatial correlations for area and topology are
weak.  Pairs of bubbles switch as the strain exceeds their local yield
points.  Although more frequent, the avalanches are small, without
long correlation lengths.  Figure 8(c) shows the power spectra for T1
events for a highly disordered structure with $\mu_2(n)=1.65$.  We
observe no power law behavior, even at high shear rates.  Thus a
highly disordered foam resembles a homogeneous but nonlinear viscous
fluid.  

Over a range of structural disorder the topological rearrangement
events are strongly correlated.  The question naturally arises whether
the transition between these correlated and uncorrelated regimes is
sharp or smooth, and what determines the transition points.  We are
currently carrying out detailed simulations involving different
structural disorder to study this transition.

Previous simulations \cite{durian97} and experiments
\cite{dennin_knobler97} measured $\overline{N}$, the average number of
T1 events per bubble per unit shear, and concluded that $\overline{N}$
was independent of the shear rate.  Our simulation results in
three different foams with shear rates covering two orders of
magnitude, however, disagree.  As shown in Fig. 9, our data indicate
that $\overline{N}$ depends sensitively on both the polydispersity of
the foam and the shear rate.  Only at large shear rates does
$\overline{N}$ seem to be independent of the shear rate, which might
correspond to the above-mentioned experiments.

The avalanches and $1/f$ power spectra resemble a number of
systems with many degrees of freedom and dissipative dynamics which
organize into marginally stable states~\cite{bak87-88}.  
Simple examples include stick-slip models, driven chains of
nonlinear oscillators, and sandpile models.  In sandpile models, both
the energy dissipation rate (total number of transport events at each
time step) and the output current (the number of sand grains leaving
the pile) show power-law scaling in their distributions.  In
particular, if the avalanches do not overlap, then the power spectrum
of the output current follows a power law with a finite size
cutoff~\cite{jensen89-91}.  The $1/f$-type power spectra result from
random superposition of individual avalanches \cite{hwa92}.  

If the analogy with sandpiles holds, we should expect the power spectra 
of the time derivative ${d\phi / dt}$, of the stored energy, 
{\it i.e.\/} the energy change at every time step, to be $1/f$-like, and
thus the power spectra of $\phi$ to be $f^{-2}$.  However, in our
simulations ${d\phi / dt}$ does not show $1/f$-type 
broadband noise.  Figures 10 (a-c) show the corresponding power
spectra for $\phi$ from the same simulations as Fig. 8, which are
obviously not $f^{-2}$, {\it i.e.\/} the topological rearrangements are
not in the same universality class as sandpiles.  In particular,
Fig. 8(c) shows a complicated trend: the power spectrum changes from a
small slope at shear rate $\beta=0.001$ to $f^{-0.8}$ spanning over 4
decades at $\beta=0.005$.  But as the shear rate increases, the power
law disappears.  Instead, a flat tail develops at high frequencies due
to Gaussian noise.  Other different slopes appear over different
regimes of different sizes, indicating the existence of multiple
time-scales and length-scales.  We will further explore the
implications of these spectra for $\phi$ elsewhere~\cite{jiang99}.

\section{Effects of structural disorder}

As structural disorder plays an important role in mechanical response,
we study the effect of disorder on the yield strain and the evolution
of disorder in foams under shear.  We define the yield strain, at
which the first T1 avalanches occur, as the displacement at the top
boundary of the foam divided by half the height of the foam (since the
zero strain is in the middle of the foam) rescaled by the average 
bubble width.

Figure 11(a) shows the yield strain as a function of shear rate
$\beta$ for different foam disorders.  We find that for an ordered
foam at low shear rates, when a sliding plane occurs in the middle of
the foam, the yield strain is independent of shear rate.  We expect
this independence because T1 events occur almost simultaneously in the
sliding plane, and the bubble size determines the yield strain.  At
high shear rates, T1 events distribute more homogeneously throughout
the foam which lowers the yield strain.  The upper limit for the yield
strain in an ordered foam is $2/\sqrt{3}$, when all the vertices in a
hexagonal bubble array simultaneously become four-fold under shear.
The nucleation of topological defects (5-and-7-sided bubble pairs) and 
their propagation in foams lower the yield strain.  But the yield
strain does not reach zero even at a very high shear rate of $\beta=
0.05$.  An ordered foam remains a solid with finite yield strain.  For
a disordered foam, the yield strain is lower for higher shear rates;
and at the same shear rate, the yield strain decreases drastically to
zero as disorder increases --- the foam changes from a viscoelastic
solid to a viscoelastic fluid.  

The most commonly used measure for topological disorder is the second
moment of the topological distribution, $\mu_2(n)$.  During
diffusional foam coarsening, the topological distribution tends to a
stationary scaling form and $\mu_2(n)$ assumes a roughly constant
value.  Experiments on soap foams with up to 10000 bubbles in the initial
state \cite{glazier89,stavans89} and early smaller simulations
\cite{weaire90} gave a value of $\mu_2(n)=1.4$ in the scaling regime.
Other simulations showed a slightly lower value of
$\mu_2=1.2$~\cite{herdtle92}.  Weaire {\it et
al.\/}~\cite{weaire_bolton92} reported shear-induced ordering, {\it i.e.\/}
reduction of $\mu_2(n)$ with shearing.  However, in our simulations,
foams with initial $\mu_2(n)$ ranging from $0.81$ to $2.02$ show no
shear-induced ordering.  Instead, $\mu_2(n)$ increases and never
decreases back to its initial unstrained value.  Figure 11(b) shows
the evolution of $\mu_2(n)$ for a variety of initial topological
distributions.  The difference between the simulations of Weaire {\it  
et al.\/}~\cite{weaire_bolton92} and ours is not surprising.  Weaire
{\it et al.\/} applied step strain and observed the resulting
equilibrated pattern.  In our simulations, bubbles are constantly
under shear, {\it i.e.\/} the foam is not in equilibrium.  The topological
disorder, as measured by $\mu_2(n)$, therefore increases as the energy 
accumulates and decreases as the energy releases, as does the number
of topological events, and does not necessarily settle to an
equilibrium value.   

In an ordered foam at shear rate $\beta=0.01$ [Fig. 4] with
separated T1 avalanches, $\mu_2(n)$ fluctuates in synchrony with the
total energy, shown in Fig. 11(c).  When the avalanches overlap,
$\mu_2(n)$ fluctuates more smoothly, but almost always has a positive
correlation with the total stored energy. 

Notice that in the energy-strain plots [Fig. 2(b), Fig.4(b)], stored
energy slowly increases over long times, because we continuously apply
shear and the foam is always out of equilibrium.  Bubbles do not fully 
recover their original shapes.  This deformation slowly accumulates at
long times.  In disordered foams, topological rearrangement may enhance
the spatial correlation of bubbles, {\it i.e.\/} small
bubbles cluster over time, as predicted by Langer and Liu's bubble
model~\cite{langer97}.  We will report results on spatial correlations
for bubble topology $n$, and area $a$ elsewhere \cite{jiang99}.

\section{Conclusions} 

We have included a driving term in the large-Q Potts model to apply
shear to foams of different disorder.  When the driving rate is too
fast for the foam to relax, the system falls out of equilibrium.
The mechanical response then lags behind the driving shear, resulting
in hysteresis.  Our model differs from most well studied
driven spin models: our spins do not couple to an external field the
way Ising spins couple to an oscillating magnetic field,
and all action occurs only at the domain boundaries.  

Because of the difficulty in characterizing local stress and strain in
Potts model foams, we have chosen to use the rescaled total bubble
wall length, $\phi$, as the order parameter for hysteresis.  While the 
hysteresis loops reflect the nonlinearity and metastability of bubble
configurations, it is still an open question whether we can find
more appropriate order parameter(s) that will provide more insight into
the dynamics of T1 events.  Another consequence of this difficulty is
the lack of a clear quantitative description of the viscosity in our
simulated foams in terms of the model parameters, a difficulty
mirrored in the lack of understanding of effective foam viscosity in
experiments.  As mentioned above, a fundamental problem is the lack of
experimental data on the viscosity of two-dimensional foams.  We hope
that these simulations will motivate new experiments in this direction.  

The local cellular patterns characteristic of T1 events in foams are
strikingly similar to the low temperature defects and the
hexatic-square Voronoi patterns observed in two-dimensional (particle)
systems, {\it e.g.\/} two-dimensional liquid crystals and colloidal
suspensions, where studies have focused on the melting phase transitions
\cite{stranburg88,denetal89}.  This similarity lead us to try the defect 
description used in melting studies, namely the
nearest-neighbor-bond-orientation order parameter
$\sum_{n}\exp(6i\theta_n)$, where $\theta_n$ is the angle between two
neighboring bonds.  However, we found it insensitive to the
orientation change of bubble walls during T1 events and thus not a 
useful order parameter.  How the nucleation and propagation of
topological defects in a sheared foam relate to the nucleation and
role of topological defects in the two-dimensional melting studies
remains an interesting question.

We have demonstrated three different hysteresis regimes in an 
ordered foam under oscillating shear.  At small strain amplitudes, bubbles
deform and recover their shapes elastically after stress release.
The macroscopic response is that of a linear elastic solid.  For larger
strain, the energy-strain curve starts to
exhibit hysteresis before any topological rearrangements occur,
indicating a macroscopic viscoelastic response.  Increasing the strain
amplitude increases the area of the hysteresis loop.  When the applied
strain amplitude exceeds a critical value, the yield strain, T1 events
occur and the foam starts to flow, and we observe macroscopic
irreversibility. 

We are currently testing this observation in an experiment similar to
\cite{hohler97}, applying periodic 
boundary shear to a homogeneous foam, and measuring the total bubble
wall length directly (instead of using diffusion wave spectroscopy) to 
obtain $\phi$.  We can directly compare this data with the predicted 
three distinct behaviors.  The viscoelasticity of foams is better
characterized using the complex modulus $G(\omega)$ \cite{sollich97},
which we plan to use in future investigations. 

The comparison between the mechanical responses of ordered and
disordered foams provides some insight into the relation between local
structure and macroscopic response.  An ordered foam has a finite
yield strain.  Structural disorder decreases the yield strain;
sufficiently high disorder changes the macroscopic response of a foam
from a viscoelastic solid to a viscoelastic fluid.  A random foam
with broad topology and area distributions lacks the linear elastic and
viscoelastic solid regimes.  Any finite stress can
lead to topological rearrangements of small bubbles and thus to
plastic yielding of the foam.  More detailed simulations and
experiments are needed to determine the dependence of the yield strain
on the area and topological distributions of the foam, and on the
shear rates.  High shear rates effectively introduce more topological
defects into the foam, as manifested in ordered foams driven at high
shear rates.  Local topological rearrangements (the appearance of
non-hexagonal bubbles) occur throughout the foam, resulting in more
homogeneous flow behavior, as in disordered foams.  

Our simulations show that $\overline{N}$, the average number of T1
events per bubble per unit shear, is sensitive to the area
distribution of the foam and the shear rate.  Only for a small range
of shear rates do foams having similar distributions show similar
values of $\overline{N}$, which may explain previous
studies~\cite{durian97,dennin_knobler97}.  Our results emphasize the
importance of both structural disorder and configurational
metastability to the behavior of soft cellular materials.    

In disordered foams, the number of T1 events is not directly
proportional to the elastic energy release, because each T1 event is
non-local and different T1 events can release different amounts of
energy.  Therefore, we count T1 events and energy release
separately. 

Avalanche-like topological rearrangements play a key role in foam
rheology.  Our simulations show that T1 events do not have finite
long-range correlations for ordered structures or at low shear rates,
consistent with experimental observations.  As the shear rate or
structural disorder increases, the topological events become more
correlated.  Over a range of disorders, the power spectra are $1/f$.
As Hwa and Kardar pointed out, $1/f$ noise may arise from a random
superposition of avalanches \cite{hwa92}.  These $1/f$ spectra suggest that 
avalanches of different sizes, although they overlap, are independent
of each other.  Both greater structural disorder and higher
shear rate introduce a flat tail at the high frequency end, a signature
of Gaussian noise, but do not change the exponent in the power law region.

However, unlike the sandpile model, the power spectra of the total
energy, rather than of the energy dissipation, show a similar trend
toward $1/f$.  One major difference between T1 avalanches and sand
avalanches is that each sand grain carries the 
same energy, while each T1 event can have a different energy.  A better 
analogy may be a ``disordered sandpile'' model, where the sand grains
have different sizes or densities, and avalanches overlap.
  
Avalanches of T1 events decrease the stored elastic energy, leading to
foam flow.  How do single T1 events contribute to the global response?
Magnetic fluid foam experiments offer a unique opportunity to locally
drive a vertex and force a single T1 event (or a T1 avalanche) by a
well-controlled local magnetic field.  We are investigating the
effects of single T1 events using magnetic fluid foam experiments and
the corresponding Potts model simulations \cite{elias98-2}.   

\newpage
\section*{Acknowledgment}
We would like to thank F. Graner, M. Sano, S. Boettcher and I. Mitkov
for fruitful discussions.  This work was supported in part by NSF
DMR-92-57001-006, ACS/PRF and NSF INT 96-03035-0C, and in part by the
US Department of Energy.     

%\newpage

\newpage

\section*{Figure Captions}

Figure 1. Schematic diagram of a T1 event, where bubbles $a, b, c$ and
$d$ swap neighbors.  Notice that as the edge between the pair of
vertices shrinks, the contact angles not in contact with this edge
remain $120^\circ$. 

\bigskip 

Figure 2. An ordered foam under boundary shear: (a) Snapshots; 
different shades of grey encode bubble topologies (lattice size
$256\times256$).  (b) Energy-strain curve and the number of T1s
presented in 50 MCS bins.  

\bigskip 

Figure 3. Energy-strain curves for ordered foam under periodic
boundary shear: 
(a) Numbers above the figures are $\gamma_0$.  Progressively
increasing shear amplitude at ${\cal J}_{ij}=3$ leads to a transition between
three types of hysteresis: $\gamma_0=1.0$ corresponds to an elastic 
response, $\gamma_0=3.5$ shows viscoelastic response (before any T1 event
occurs), $\gamma_0=7.0$ is a typical response when only one T1 event occurs
during one cycle of strain loading.  The intermediate steps show that
the transition between these three types is smooth. 
(b) Numbers above the figures are ${\cal J}_{ij}$.  Progressively decreasing
liquid viscosity (increasing ${\cal J}_{ij}$ at $\gamma_0=7$) shows a similar
transition between elastic (${\cal J}_{ij}=10$) and viscoelastic
(${\cal J}_{ij}=5$)
regimes, and flow due to T1 events (${\cal J}_{ij}=3$ for one T1 event and
${\cal J}_{ij}=1$ for three T1 events during one strain cycle. 
(c) Phase diagram of hysteresis in the parameter space $\gamma_0$
{\it vs.\/} ${\cal J}_{ij}$; 
(d) Effect of progressively increasing temperature $T$ (${\cal J}_{ij}=3$,
$\gamma_0=4$).  
All data shown here are averaged over 10 periods.  
 
\bigskip 

Figure 4. An ordered foam under bulk shear with shear rate $\beta = 0.01$: 
(a) Snapshots; shades of grey encode bubble topologies as in Fig. 2. 
(b) Energy-strain curve and the number of T1s.  The magnified view in 
the box shows the correlation between stress releases and overlapping 
avalanches of T1 events. 

\bigskip

Figure 5. An ordered foam under bulk shear with shear rate $\beta = 0.05$:
(a) Snapshots; (b) Energy-strain curve and the number of T1s.  
The magnified view in the box shows the correlation between stress
releases and overlapping avalanches of T1 events. 

\bigskip

Figure 6. A disordered foam under boundary shear: (a) Snapshots;
shades of grey encode bubble topologies (lattice size $256\times256$). 
(b) Energy-strain curve and number of T1 events presented in 100 MCS bins. 

\bigskip

Figure 7. A disordered foam under bulk shear at shear rate $\beta=0.01$: 
(a) Snapshots; shades of grey encode bubble topologies (lattice size
$256\times256$):  (b) Energy-strain curve and number of T1 events.

\bigskip

Figure 8. Power spectra of the number of T1 events: (a) An ordered foam for
three shear rates $\beta=0.01$, $0.02$ and $0.05$ respectively; 
(b) A disordered foam [$\mu_2(n)=0.81, \mu_2(a)= 7.25$] for five shear
rates from $0.001$ to $0.05$; 
(c) A very disordered foam [$\mu_2(n)=1.65,\mu_2(a)= 21.33$] for three
shear rates.

\bigskip

Figure 9. Number of T1 events per unit shear per bubble as a function of
shear rate for four foams: squares correspond to a foam of 180
bubbles, with $\mu_2(n)=1.65$, $\mu_2(a)=21.33 $; stars correspond to
a foam of 246 bubbles with $\mu_2(n)=1.72$, $\mu_2(a)=15.1$; triangles
correspond to a foam of 377 bubbles, with $\mu_2(n)=1.07$,
$\mu_2(a)=2.50$; and circles correspond to a foam of 380 bubbles, 
with $\mu_2(n)=0.95$, $\mu_2(a)= 2.35$.  The inset shows on a log-log
scale that $\overline N$ varies by several orders of magnitude. 

\bigskip
Figure 10. Power spectra of the energy: (a) An ordered foam for three
shear rates.
(b) A disordered foam [$\mu_2(n)=0.81,\mu_2(a)= 7.25$] for five shear
rates.
(c) A very disordered foam [$\mu_2(N)=1.65, \mu_2(a)= 21.33$] for five
shear rates. 

\bigskip

Figure 11. (a) Yield strain as a function of shear rate.
(b) Evolution of $\mu_2(n)$ under constant bulk shear, legend denotes
the initial $\mu_2(n)$.
(c) Evolution of $\mu_2(n)$ under steady bulk shear for an ordered
foam, showing the correlation between the stress decreases and $\mu_2(n)$;


\begin{references}{}
\bibitem{note1} Liquid can be included as a special ``phase'' of spin;
  the coupling strengths now depend on the gas and liquid phases, see
  \cite{jiang96}.
\end{references}
\end{document}